\documentclass[12pt]{article}
\usepackage{epsfig}
\begin{document}

\begin{center}
{\Large\bf
Cosmic Ray Backgrounds in an LBNE Far Detector on the surface}
\\ 
{\bf
Lisa Goodenough, Maury Goodman, Jon Paley\\
{\it Argonne National Lab}\\
John Learned \\
{\it University of Hawaii} 
\\ Gavin Davies, Mayly Sanchez\\
{\it Iowa State University}
\\
14 June 2012
}
\end{center}
\normalsize
\begin{abstract} A surface Liquid Argon Far Detector for LBNE has formidable
background issues from cosmic rays.  There is no evidence they can be overcome.
\end{abstract}
\section{Introduction}
\par The far detector technology for LBNE is a liquid Argon TPC.
For cost reasons, a surface detector is being considered
by the LBNE reconfiguration process.
For a repetition rate of 1.33 seconds for 1.7 $10^7$ s per year, 
there are 1.31 $10^7$ spills per year of 10 microseconds each
or 131 seconds of live time per year.  With a surface cosmic
ray rate of approximately 100 $\mu/(m^2s)$, and a surface
area 24 m $\times$ 49 m, there are 15M $\mu$ per year
entering through the top plus approximately another 15\%
entering the sides, or 18M per year total in the 131 s.  

\par With a drift time of 4 ms, there are thus 7.1B muons
during drift times associated with beam spills.\footnote{We assume
there are no photon counters.  With 100\% efficient photon counters,
the effective overlap time would approach the beam spill length.}
In contrast,
the signal is 170 events per year for a 34 kT detector at Ash River.

\par A previous note looked at the qualitative issues
associated with a liquid Argon Far Detector on the surface.\cite{bib:pordes}
It identified four possible issues related to cosmic rays:  {\it i.)
 they could generate so much data that the DAQ is overwhelmed, 
ii.)  they might obscure such a large fraction of the volume of
the detector that they overlap the events of interest to the point
where the events cannot be reconstructed accurately,
iii.) they might overwhelm the reconstruction and analysis such
that the computing time required simply to remove the cosmic rays
from the analysis is prohibitive,
iv.)  they could generate interactions in the argon which mimic
the neutrino events of interest.}
We only consider item (iv) and do not address the others in this note.
There is an intuitive feeling among many that this will not be a
problem despite the huge muon rate.  At near detectors or short-baseline
detectors, muons have
not been a problem.  However the requirements on signal to background
are vastly different due to the larger signal rates at near detectors.
A different argument can be found in Reference 1, ``It is indeed
highly unlikely that any muon with sufficient energy to produce
a background event will be missed."  We question that statement
for two reasons:
\begin{enumerate}
\item It is possible for muons to miss the active part
of the detector, but to make
other neutral particles which enter the detector and mimic
a background event.  This is particularly true for a modular
detector with gaps and/or dead areas.
\item It is possible to see the muon and to fail to have enough
information to unambiguously reject the background event.  Some showers
near muons (from bremsstrahlung , decay-in-flight, neutral hadrons from spallation,
...) will need to be cut.
\end{enumerate}
To fully determine these backgrounds, we either need sophisticated
Monte Carlos or surface background measurements in an identical
detector, together with full pattern recognition algorithms for 
the signal electron neutrinos.  Because cosmic ray muons near the 
surface tend to be associated with showers, there can be pathological 
events with complex topologies which are hard to model. 

\par One simple calculation that we can do is to find the
efficiency for retaining signal $\nu_e$ events if we have to make a cut as a
function of distance of closest approach between a $\nu_e$
vertex and the closest cosmic ray muon.  (This is one of many
possible ways to remove $\mu$ induced background.)
We imagine a cylinder around the muon
with a radius r.   We might choose to cut any event, for example,
within 3 radiation lengths of the muon.  Our cosmic ray muon
rate can be written:
\begin{equation}
100 \mu (m^2 s)^{-1} = 1 \mu [\pi (90~cm)^2\times(4 ms)]^{-1}
\label{eq:1}
\end{equation}
which implies that on average, there is a muon somewhere
within a circle with a radius of 90 cm from every signal 
event.  If we make a distance cut r from the vertex,
the efficiency is
\begin{equation}
\epsilon \sim 1-(\pi r^2)/(\pi (90~cm)^2
\label{eq:2}
\end{equation}

If we cut within 9 cm of each muon, this is 99\% efficient.
If we have to cut within 3 radiation lengths (3 $\times$ 14 cm) the
efficiency would be  78\%.  If we choose 2 collision lengths,
(2 $\times$ 55 cm), the efficiency is effectively zero.

\section{Discussion and Summary}

The background from these drifting muons is reduced as the
detector is made modular and the drift distance decreases.  But
this increases the background from unseen muons.

Any backgrounds associated with the muons are likely strongly
falling functions of shower energy.  This makes them relatively
a larger problem at Homestake where power for mass hierarchy and CP tests comes
from lower energy electrons.

In NOvA, the pattern recognition for electron
showers is not as powerful as in a liquid Argon detector.  However
the background starts out a factor of 400 lower, due to the drift
time.  The radius in equation \ref{eq:1} for the 10 $\mu$s spill is 18 m.
There is a plan to measure potential cosmic ray backgrounds
in NOvA using the Near Detector on the Surface.\cite{bib:4647}

We conclude that it is not clear whether a Liquid Argon TPC on the surface
for the beam physics goals of LBNE would have acceptable or
unacceptable backgrounds.  There are reasons for concern which have
not been addressed.

\hfil\break
\end{document}